\documentclass[twocolumn,showpacs,showkeys,superscriptaddress]{revtex4}

\usepackage{amsmath}
\usepackage{amssymb}
\usepackage{graphicx}
\usepackage{color}
\usepackage{verbatim}

\begin{document}

{\raggedleft {\it Published in Physics of Plasmas 23, 032111 (2016)}}
\\

\title{Visco-Resistive Plasmoid Instability}
\author{Luca Comisso}
\email{lcomisso@princeton.edu}
\affiliation{Department of Astrophysical Sciences and Princeton Plasma Physics Laboratory, Princeton University, Princeton, New Jersey 08544, USA}
\affiliation{Istituto dei Sistemi Complessi - CNR and Dipartimento Energia, \\ Politecnico di Torino, Torino 10129, Italy}
\author{Daniela Grasso}
\affiliation{Istituto dei Sistemi Complessi - CNR and Dipartimento Energia, \\ Politecnico di Torino, Torino 10129, Italy}

\begin{abstract}
The plasmoid instability in visco-resistive current sheets is analyzed in both the linear and nonlinear regimes. 
The linear growth rate and the wavenumber are found to scale as $S^{1/4} {\left( {1 + {P_m}} \right)}^{-5/8}$ and $S^{3/8} {\left( {1 + {P_m}} \right)}^{-3/16}$ with respect to the Lundquist number $S$ and the magnetic Prandtl number $P_m$. Furthermore, the linear layer width is shown to scale as $S^{-1/8} {(1+P_m)}^{1/16}$. 
The growth of the plasmoids slows down from an exponential growth to an algebraic growth when they enter into the nonlinear regime. In particular, the time-scale of the nonlinear growth of the plasmoids is found to be $\tau_{NL} \sim S^{-3/16} {(1 + P_m)^{19/32}}{\tau _{A,L}}$. The nonlinear growth of the plasmoids is radically different from the linear one and it is shown to be essential to understand the global current sheet disruption. 
It is also discussed how the plasmoid instability enables fast magnetic reconnection in visco-resistive plasmas.
In particular, it is shown that the recursive plasmoid formation can trigger a collisionless reconnection regime if $S \gtrsim  L_{cs} {(\epsilon_c l_k)^{-1}} {(1 + {P_m})^{1/2}}$, where $L_{cs}$ is the half-length of the global current sheet and $l_k$ is the relevant kinetic length scale. On the other hand, if the current sheet remains in the collisional regime, the global (time-averaged) reconnection rate is shown to be $\left\langle {{{\left. {d\psi /dt} \right|}_X}} \right\rangle  \approx  \epsilon_c v_{A,u} B_{u} {(1 + {P_m})^{-1/2}}$, where 
$\epsilon_c$ is the critical inverse aspect ratio of the current sheet, while $v_{A,u}$ and $B_{u}$ are the Alfv\'en speed and the magnetic field upstream of the global reconnection layer.
\end{abstract}



\maketitle

\section{Introduction}

During the last few years there have been great advances in understanding the nature and role of the plasmoid instability \cite{Lou_2016}, i.e., the instability of reconnecting current sheets leading to the formation of secondary magnetic islands (plasmoids).
Numerical simulations providing clear indications that thin reconnecting current sheets may be unstable to the formation of secondary islands date back at least to the mid-eighties \cite{Steinolfson_1984,Matt_Lam_1985,Bisk_1986,Lee_Fu_1986}, but it has been only after a seminal paper by Loureiro and coworkers \cite{Lou_2007} that the scientific community has begun to show a distinct interest in the plasmoid instability and its effects on magnetic reconnection. 
In their paper, Loureiro {\it et al.} \cite{Lou_2007} presented a linear theory of the plasmoid instability in Sweet-Parker current sheets, showing that the growth rate and the wavenumber of the instability scale as $\gamma_{\max} L_{cs}/v_{A,u} \sim S^{1/4}$ and $k_{\max } L_{cs} \sim S^{3/8}$, respectively. Here, $S$ is the Lundquist number based on the half-length of the current sheet ($L_{cs}$) and the Alfv\'en speed upstream of the sheet ($v_{A,u}$). This implies that the linear growth of the plasmoids is surprisingly fast for large values of the Lundquist number.

The effects of the plasmoid instability on the magnetic reconnection rate became clear shortly after Loureiro's paper, when Bhattacharjee and coworkers \cite{BHYR_2009} showed that the predictions of the Sweet-Parker model \cite{Biskamp_2000} break down for high-Lundquist-numbers as a consequence of the plasmoid instability. The Sweet-Parker model gives a reconnection rate $\sim S^{-1/2} v_{A,u} B_{u}$, where $B_{u}$ is the upstream magnetic field, but Bhattacharjee {\it et al.} \cite{BHYR_2009} showed that this cannot be valid for arbitrarily large $S$-values. In particular, they found from numerical simulations that the reconnection rate becomes nearly independent on the Lundquist number when the current sheet exceeds a critical Lundquist number $S_c$. Note that the reconnection rate calculated by Bhattacharjee and coworkers was averaged over time, since in the high-Lundquist-number regime the reconnection process is strongly time-dependent due to the continuous formation and ejection of plasmoids \cite{BHYR_2009,Daugh_2009,Cassak_2009}. An estimation of the time-averaged reconnection rate in this regime was proposed by Huang and Bhattacharjee \cite{HB_2010} as well as by Uzdensky and coworkers \cite{ULS_2010}. 
They showed that, as a result of the plasmoid instability, the time-averaged reconnection rate in statistical steady-state is $\sim 0.01 v_{A,u} B_{u}$, independent of the Lundquist number and much higher than the Sweet-Parker rate for very large $S$-values. 
These crucial implications of the plasmoid instability caused a rethinking of the traditional magnetic reconnection theory, with a particular impetus for the investigation of the linear properties of this instability  \cite{SLUSC_2009,Ni_2010,Baalrud_2011,Baalrud_2012,LSU_2013,Huang_2013,Pucci_2014,Tenerani_2015,Nemati_2015} and its effects on the reconnection rate \cite{SC_2010,HBS_2011,Loureiro2012,Ni_2012,HB_2013,Murphy_2013,CGW_2014,Yu_2014,Ni_2015,Comisso2015,Ebrahimi2015} and particle acceleration \cite{Sironi2014,Guo2014,Dahlin2014,Guo2015,Li2015}.

In the present paper we have three main objectives. First, we intend to generalize the linear analysis presented by Loureiro and coworkers \cite{Lou_2007} to non-negligible values of the plasma viscosity. Then, we want to extend the analysis of the plasmoid instability also to the nonlinear regime. Finally, we aim to estimate the effects of this instability on the global reconnection rate. We point out that for all these issues we will consider arbitrary values of the magnetic Prandtl number.

The linear analysis will show that the plasmoid instability remains fast also in the general case of non-negligible plasma viscosity. This means that its rapidity is a robust feature in the linear regime. However, we must recall that the linear layer width has to be much smaller than the current sheet width for the linear analysis to apply. Thus, one has to consider also the nonlinear regime to evaluate the overall growth rate of the plasmoid instability.
We will see that the growth of the plasmoids slows down when their width becomes comparable or larger than the linear layer width. In particular, we will show how the time-scale of the plasmoid nonlinear growth depends on the Lundquist number and the magnatic Prandtl number. It will also be shown that the recursive plasmoid formation (due to the instability of the secondary current sheets) may result in a speed up of the global current sheet disruption. 
The break up of the global current sheet has the effect of accelerating the magnetic reconnection process with respect to the case of a current sheet that would remain stable for arbitrary values of the Lundquist number. We will show that for high-Lundquist-numbers the (time-averaged) reconnection rate becomes independent of the Lundquist number but not on the magnetic Prandtl number. In particular, the reconnection rate decreases for increasing values of the magnetic Prandtl number. 
Finally, we will see that plasma viscosity has also the effect of increasing the value of the global Lundquist number required to trigger collisionless reconnection.

\section{Model Equations and Equilibrium}  \label{sec2}

%-------MODEL EQUATIONS-------

We consider the two-dimensional ($\partial_z = 0$) incompressible MHD equations, which can be conveniently written in terms of the magnetic flux function $\psi (x,y,t)$ and the stream function $\phi (x,y,t)$ as \cite{Biskamp_2000}
\begin{equation} 
{\partial_t}\psi  + {\mathbf{v}} \cdot \nabla \psi   =  - \eta j + {E_0} \, , \label{e1}
\end{equation}
\begin{equation} 
\partial_t \omega + {\mathbf{v}} \cdot \nabla \omega   =  {\bf{B}} \cdot \nabla {j} + \nu {\nabla ^2}{\omega} \, , \label{e2}
\end{equation}
where
\begin{equation} 
{\bf{B}} = \nabla \psi \times {{\bf{e}}_z} \, , \quad {\bf{v}} = {{\bf{e}}_z} \times \nabla \phi \, 
\end{equation}
are the magnetic and velocity fields, while
\begin{equation} 
j = - \nabla^2 \psi \, , \quad \omega = \nabla^2 \phi \, 
\end{equation}
are the electric current density and plasma vorticity, both in the $z$ direction. Furthermore, $E_0$ is a constant of integration representing the equilibrium electric field. The mass density is assumed to be uniform and is normalized such that $\rho =1$. Lengths are normalized to a convenient scale length $L$, magnetic field to a convenient scale field strength $B_0$, and time to the Alfv\'en time $\tau_A = L/v_A$, where $v_A$ is the Alfv\'en speed based on $B_0$. In this normalization, the resistivity  $\eta$ and the viscosity $\nu$ are reciprocals of the Lundquist number and the (kinetic) Reynolds number, respectively. Therefore, the ratio $P_m = \nu / \eta$ indicates the magnetic Prandtl number.

%-------EQUILIBRIUM-------

We consider the equilibrium configuration employed by Loureiro {\it et al.} \cite{Lou_2007} in order to investigate the stability of a Sweet-Parker type current sheet. As pointed out in their landmark paper, this configuration is not intended to retain all the features of such a sheet, but only those which may be regarded as the most important for its stability; namely, a linearly increasing velocity in the ouflow direction and a sheared magnetic field in the inflow direction.

Assuming an equilibrium flow profile 
${{\mathbf{v}}_0} = \left( { - {\Gamma _0}x,{\Gamma _0}y} \right)$ for $\left| x \right| \leqslant x_0$ (inside the current sheet), 
${{\mathbf{v}}_0} = \left( { - {\Gamma_0}x_0 ,0} \right)$ for $x \geqslant x_0$, and 
${{\mathbf{v}}_0} = \left( { {\Gamma_0}x_0 ,0} \right)$ for $x \leqslant -x_0$, the equilibrium stream function turns out to be
\begin{equation}  \label{}
{\phi_0} (x,y) =
\begin{cases}
    + {\Gamma_0} x y \, ,    &    \left| x \right| \leqslant x_0 \, , \\
    + {\Gamma_0} x_0 y \, ,   &    x \geqslant x_0 \, , \\
    - {\Gamma_0} x_0 y \, ,    &    x \leqslant  - x_0 \, .
\end{cases}
\end{equation}
Here, ${\Gamma_0} \equiv {v_d}/{L_{cs}}$ is a constant velocity gradient, with $v_d=v_{0y}(0,L_{cs})$ and $L_{cs}$ indicating the downstream velocity and the half-length of the current sheet, respectively. In the presence of non-negligible viscosity effects, the force balance along the current sheet gives that \cite{Park_1984} 
\begin{equation} 
v_d = v_{A,u} {\left( {1 + P_m} \right)^{-1/2}} \, ,
\end{equation}
where $v_{A,u}$ is the Alfv\'en speed upstream of the sheet. The downstream velocity decreases as $v_d \approx v_{A,u} P_m^{-1/2}$ for $P_m \gg 1$, while the standard Sweet-Parker result \cite{Biskamp_2000} $v_d \approx v_{A,u}$ is recovered in the limit $P_m  \ll 1$. Also, mass conservation for an incompressible flow implies that the upstream velocity is related to the downstream velocity and the aspect ratio of the current sheet as $v_u = v_{0x}(\delta_{cs},0) = - v_d {\delta_{cs}} /{L_{cs}}$, where \cite{Park_1984}
\begin{equation} \label{delta_cs}
\delta_{cs} = {\left( {\eta \frac{{{L_{cs}}}}{{{v_d}}}} \right)^{1/2}} = {L_{cs}}\frac{{{{\left( {1 + {P_m}} \right)}^{1/4}}}}{{{S^{1/2}}}}  \, 
\end{equation}
is the characteristic half-width of the current sheet. Here, $S \equiv {L_{cs}}{v_{A,u}}/\eta $ is the Lundquist number based on the current sheet length and the upstream Alfv\'en speed. Again, the standard Sweet-Parker width \cite{Biskamp_2000} ${\delta _{cs}} \approx {L_{cs}}{S^{-1/2}}$ is recovered in the limit $P_m  \ll 1$, while the opposite limit $P_m \gg 1$ yields $\delta_{cs} \approx L_{cs} S^{-1/2} P_m^{1/4}$.

The equilibrium magnetic field, which is assumed to be in the form ${{\mathbf{B}}_0} = \left( {0,{B_{0y}}(x)} \right)$, can be determined by solving Eq. (\ref{e1}) with $\partial_t = 0$ and the equilibrium flow profiles already specified. In this case, Eq. (\ref{e1}) reduces to
\begin{equation} \label{Equation_eq}
\delta_{cs}^2 \frac{{d{B_{0y}}}}{{dx}} + x{B_{0y}} = \frac{{{E_0}}}{{{\Gamma _0}}} \, 
\end{equation}
for $\left| x \right| \leqslant x_0$. Introducing the new variable
\begin{equation} 
\xi \equiv \frac{x}{\delta_{cs}} \, 
\end{equation}
and using ${B_{0y}}(0)=0$ as boundary condition, the solution of Eq. (\ref{Equation_eq}) is 
\begin{equation}  \label{Sol_Loureiro_eq}
{B_{0y}}\left( {\,\xi } \right) = \alpha \, {e^{ - {\xi ^2}/2}}\sqrt {\frac{\pi }{2}} \operatorname{erfi}\left( {\frac{\xi }{{\sqrt 2}}} \right) \, , 
\end{equation}
where 
\begin{equation} 
\operatorname{erfi} \left( x \right) \equiv  - i\operatorname{erf} \left( {ix} \right) = \frac{2}{{\sqrt \pi  }}\int_0^x {{e^{{z^2}}}dz}  \, 
\end{equation}
is the imaginary error function and $\alpha = E_0 / \Gamma_0 \delta_{cs}$. Therefore, since for $\left| x \right| \geqslant x_0$ (outside the current sheet) the magnetic field is constant by virtue of Eq. (\ref{e1}) and the specified equilibrium flow profiles, we can write {\it Loureiro's 1-D equilibrium} \cite{Lou_2007} as
\begin{equation}  \label{Loureiro_eq}
B_{0y}\left( \xi  \right) =
\begin{cases}
    \alpha \, {e^{-{{\xi}^2}/2}} \displaystyle \int_{0}^{\xi} {{e^{{z^2/2}}}dz} \, ,   &   \left| \xi \right| \leqslant \xi_0 \, , \\
    + 1 \, ,   &    \xi \geqslant \xi_0 \, , \\
    - 1 \, ,    &    \xi \leqslant - \xi_0 \, .
\end{cases}
\end{equation}
Here, $\xi_0 \equiv x_0 / \delta_{cs}$ and the normalizing magnetic field strength is chosen to be the equilibrium magnetic field at $\xi = \xi_0$. The value of the constant $\alpha$ can be specified by matching the magnetic field inside the current sheet ($\left| \xi \right| \leqslant \xi_0$) with the magnetic field outside the current sheet ($\left| \xi \right| \geqslant \xi_0$). Logical matching points $\pm \xi_0$ are where the solution of Eq. (\ref{Equation_eq}) has its maximum and its minimum, i.e., where the current density vanishes. These points correspond to $\pm \xi_0 = \pm 1.307$. Therefore, requiring $B_{0y} (\pm \xi_0) = \pm 1$ yields $\alpha = 1.307$.

\section{Linear Theory} \label{sec3}

We now show that the presented equilibrium is prone to the visco-resistive plasmoid instability. To this purpose, differently from Ref. \cite{Lou_2007}, we perform a boundary layer theory that takes into account both plasma resistivity and viscosity.

\subsection{Linearized equations} 

We start by linearizing Eqs. (\ref{e1}) and (\ref{e2}) according to
\begin{equation} 
\psi (x,y,t) = {\psi_0}(x) + {\psi_1}(x,t) \, {e^{ik(t)y}} \, ,
\end{equation}
\begin{equation} 
\phi (x,y,t) = {\phi_0}(x,y) + {\phi_1}(x,t) \, {e^{ik(t)y}} \, ,
\end{equation}
where $\psi_1$ and $\phi_1$ are small perturbations to the equilibrium. As in Ref. \cite{Lou_2007}, the wavenumber along the $y$ direction is supposed to depend only on $t$, which is a valid assumption if $k$ varies sufficiently slowly across the current sheet \cite{Ni_2010}. Therefore, inside the current sheet ($\left| x \right| \leqslant x_0$) the perturbations $\psi_1$ and $\phi_1$ satisfy the equations
\begin{eqnarray}  \label{e1A}
 && {\partial_t}{\psi_1} + iy\left( {{\partial_t}k + {\Gamma_0}k} \right){\psi_1} + ik {B_{0y}} {\phi_1} - {\Gamma_0}x{\partial _x}{\psi_1} \nonumber\\
 &&\qquad = \eta \left( {\partial _x^2 - {k^2}} \right){\psi_1} \,
\end{eqnarray}
and
\begin{eqnarray}  \label{e2A}
 && {\partial_t}\left( {\partial_x^2 - {k^2}} \right){\phi_1} + iy\left( {{\partial_t}k + {\Gamma_0}k} \right) \left( {\partial_x^2 - {k^2}} \right) {\phi_1}  \nonumber\\
 && \; \; - 2k\left( {{\partial_t}k} \right){\phi_1} - {\Gamma_0}x{\partial_x}\left( {\partial_x^2 - {k^2}} \right){\phi_1} = ik \frac{{{d^2}{B_{0y}}}}{d{x^2}} {\psi_1} \nonumber\\
 && \; \; - ik B_{0y} \left( {\partial_x^2 - k^2} \right) {\psi_1} + \nu \left( {\partial_x^4 - 2{k^2}\partial_x^2 + {k^4}} \right){\phi_1} \, .
\end{eqnarray}
These equations depend on both $x$ and $y$. However, the $y$-dependence can be removed by choosing $k(t)$ to satisfy ${\partial_t} k + {\Gamma_0} k = 0$ inside the current sheet. This gives that $k(t) = {k_0}{e^{ - {\Gamma_0}t}}$ \cite{Lou_2007}, and the above equations reduce to
\begin{equation}  \label{e1AA}
{\partial_t}{\psi_1} + ik {B_{0y}} {\phi_1} - {\Gamma_0}x{\partial _x}{\psi_1} = \eta \left( {\partial _x^2 - {k^2}} \right){\psi_1} \, , 
\end{equation}  
\begin{eqnarray}  \label{e2AA}
 && {\partial_t}\left( {\partial_x^2 - {k^2}} \right){\phi_1} + 2{\Gamma_0}{k^2}{\phi_1} - {\Gamma_0}x{\partial_x}\left( {\partial_x^2 - {k^2}} \right){\phi_1} \nonumber\\
 && \qquad = ik\left[ { \frac{{{d^2}{B_{0y}}}}{d{x^2}} - {B_{0y}}\left( {\partial _x^2 - {k^2}} \right) } \right]{\psi_1}  \nonumber\\
 && \qquad + \, \nu \left( {\partial_x^4 - 2{k^2}\partial_x^2 + {k^4}} \right){\phi_1} \, .
\end{eqnarray}
Actually, in our analysis we can ignore also the time-dependence of $k$ since we look for fast growing modes \cite{Lou_2007,Ni_2010}. Alternatively, we could have started directly with a time-independent $k$, but we have adopted this approach to illustrate the domain of validity of the const-$k$ approximation. In particular, we search for solutions of the form 
\begin{equation}
{\psi_1}(x,t) = \Psi(x) {e^{\gamma t}} \, , \quad {\phi_1}(x,t) = - i \Phi(x) {e^{\gamma t}} \, ,
\end{equation}
with growth rate 
\begin{equation}
\gamma \gg \frac{v_d}{L_{cs}} \equiv \Gamma_0 \, .
\end{equation}
Indeed, in order for a tearing mode to grow it is necessary that its growth rate exceeds the shearing rate \cite{Biskamp_2000,Bulanov1979}. In this limit, $k(t) \approx k_0$ and the terms proportional to $\Gamma_0$ can be neglected. Therefore, the linear shear flow contributes only via the equilibrium profile ${B_{0y}}(x)$ \cite{Lou_2007}. Under these approximations, and with the change of variable $\xi \equiv x / \delta_{cs}$, the equations that govern the perturbed quantities $\Psi (\xi)$ and $\Phi (\xi)$ are
\begin{equation} \label{eq7}
\lambda \Psi + B_{0y} \Phi = \frac{1}{{\kappa {{(1 + {P_m})}^{1/2}}}} \left( {\frac{{{d^2}}}{{d{\xi ^2}}} - {\kappa ^2}{\epsilon ^2}} \right)\Psi \, , 
\end{equation}
\begin{eqnarray}  \label{eq8}
 && \lambda \left( {\frac{{{d^2}}}{{d{\xi ^2}}} - {\kappa ^2}{\epsilon ^2}} \right)\Phi  =  B_{0y} \left( {\frac{{{d^2}}}{{d{\xi ^2}}} - {\kappa ^2}{\epsilon ^2}} \right)\Psi - \frac{{d^2} B_{0y}}{{d{\xi ^2}}}\Psi  \nonumber\\
 && \quad \; \; \; +  \frac{{{P_m}}}{{\kappa {{(1 + {P_m})}^{1/2}}}} \left( {\frac{{{d^4}}}{{d{\xi ^4}}} - 2{\kappa ^2}{\epsilon ^2}\frac{{{d^2}}}{{d{\xi ^2}}} + {\kappa ^4}{\epsilon ^4}} \right)\Phi ,
\end{eqnarray}
where we have introduced the parameters
\begin{equation}
\lambda \equiv \frac{\gamma}{k_0 v_{A,u}} \, , \quad \epsilon \equiv \frac{\delta_{cs}}{L_{cs}}  \, , \quad \kappa \equiv k_0 L_{cs}  \, .
\end{equation}

We recall that $\epsilon \ll 1$, since we are considering highly elongated current sheets. Furthermore, we assume 
\begin{equation}
\lambda \ll 1 \, , \quad  \kappa \gg 1  \, , \quad  \kappa \epsilon \ll 1  \, ,
\end{equation}
but also 
\begin{equation}
\lambda \gg \kappa^{-1} {(1 + {P_m})^{-1/2}}  \, ,
\end{equation}
since we are looking for perturbations such that $\gamma \gg \Gamma_0$. All these assumptions are satisfied by the fastest growing mode, as can be verified {\it a posteriori}. Then, as in tearing mode theory, the problem can be made more tractable by dividing the spatial domain into two regions: an ``outer region'' ($\left| \xi \right| \gtrsim 1$), where non-ideal effects and plasma inertia are negligible, and a thin ``inner region'' ($\left| \xi \right| \ll 1$) centered on the resonant surface at $x=0$, where resistivity, viscosity, and plasma inertia can be important.

\subsection{Outer region}

In the region $\left| \xi \right| \sim 1$, neglecting plasma inertia and non-ideal effects, Eq. (\ref{eq7}) reduces to 
\begin{equation} \label{outer_eq1}
\Phi  =  - \frac{\lambda }{{{B_{0y}}}}\Psi   \, 
\end{equation}
while Eq. (\ref{eq8}) becomes
\begin{equation} \label{outer_eq2}
\frac{{{d^2}\Psi }}{{d{\xi ^2}}} = \left( {\frac{{{{{d^2}{B_{0y}}} \mathord{\left/
 {\vphantom {{{d^2}{B_{0y}}} {d{\xi ^2}}}} \right.
 \kern-\nulldelimiterspace} {d{\xi ^2}}}}}{{{B_{0y}}}} + {\kappa ^2}{\epsilon ^2}} \right)\Psi  \, .
\end{equation}
Eq. (\ref{outer_eq2}) can be solved perturbatively by exploiting the smallness of $\kappa^2 \epsilon^2$, as it has been done by Loureiro {\it et al.} \cite{Lou_2007}. Its solution can be written as
\begin{equation} \label{Loureiro_sol_1}
{\Psi ^ \pm }\left( \xi  \right) = C_1^ \pm B_{0y}\left( \xi  \right) + C_2^ \pm B_{0y}\left( \xi  \right)\int_{{\pm \xi_0}}^\xi  {{B_{0y}^{ - 2}}(z)dz} \, ,
\end{equation}
where $\pm$ refers to the solution at $\xi \gtrless 0$. The constants of integration $C_2^ \pm$ can be found by requiring the solution to be continuous across the rational surface $\xi = 0$ and adopting the approximation $B_{0y}(\xi) \approx \alpha \xi$ for small $\xi$. This gives that $C_2^+ = C_2^- = - \alpha \Psi(0)$. 
The integration constants $C_1^ \pm$ can instead be found by matching the solution (\ref{Loureiro_sol_1}) with the solution of Eq. (\ref{outer_eq2}) in the region $\left| \xi \right| \geqslant \xi_0$, which is ${\Psi ^ \pm }\left( \xi  \right) = C_3^\pm \, {e^{ \mp \kappa \epsilon \xi }}$. Matching this solution and its first derivative with the solution (\ref{Loureiro_sol_1}) and its first derivative at $\xi = \pm \xi_0$ yields $C_1^ \pm  =  \pm \alpha \Psi (0)/\kappa \epsilon$ and $C_3^ \pm  = \left( {\alpha \Psi (0)/\kappa \epsilon } \right){e^{\kappa \epsilon {\xi_0}}}$. Therefore, the ideal-MHD magnetic flux eigenfunction for Loureiro's 1-D equilibrium is \cite{Lou_2007} 
\begin{equation}  \label{Loureiro_solution}
{\Psi^\pm } ( \xi ) =
\begin{cases}
    \dfrac{\alpha }{{\kappa \epsilon}} \Psi(0) B_{0y}(\xi) \Bigg( { \pm 1 - \kappa \epsilon \displaystyle 
    \int\limits_{\pm \xi_0}^{\xi}  {{B_{0y}^{-2}}(z)dz} } \Bigg)  ,   &   \left| \xi \right| \leqslant \xi_0 , \\
    \dfrac{\alpha }{{\kappa \epsilon}} \Psi(0) \, {e^{\kappa \epsilon \left( {{\xi_0} \mp \xi } \right)}}  ,    &   \left| \xi \right| \geqslant \xi_0 .
\end{cases}
\end{equation}
This solution has a discontinuous first derivative at the rational surface $\xi = 0$. In particular, the jump in its first derivative gives the tearing stability parameter \cite{Lou_2007,FKR_1963}
\begin{equation} \label{Tearing_parameter}
\Delta ' \equiv \frac{1}{{\Psi (0)}}\left( {{{\left. {\frac{{d\Psi }}{{d\xi }}} \right|}_{{0^ + }}} - {{\left. {\frac{{d\Psi }}{{d\xi }}} \right|}_{{0^ - }}}} \right) \approx \frac{{2{\alpha ^2}}}{{\kappa \epsilon }} \, .
\end{equation}

In the following, as in the standard tearing mode theory, we will perform a matching between the solutions in the outer and inner regions by means of the parameter $\Delta '$. Note that while the expression (\ref{Tearing_parameter}) is the same as in Loureiro {\it et al.} \cite{Lou_2007}, here the inverse aspect ratio of the current sheet $\epsilon$ depends not only on the Lundquist number $S$ but also on the magnetic Prandtl number $P_m$.

\subsection{Inner region}  

In the region $\left| \xi \right| \ll 1$ we can assume ${\kappa^2}{\epsilon^2} \ll {d^2}/d{\xi^2}$ and $B_{0y} \approx \alpha \xi$. Therefore, Eqs. (\ref{eq7}) and (\ref{eq8}) reduce to
\begin{equation} \label{inner_eq1}
\lambda \Psi  + \alpha \xi \Phi  = \frac{1}{{\kappa {{(1 + {P_m})}^{1/2}}}}\frac{{{d^2}\Psi }}{{d{\xi ^2}}}  \, , 
\end{equation}
\begin{equation}  \label{inner_eq2}
\lambda \frac{{{d^2}\Phi }}{{d{\xi ^2}}} = \alpha \xi \frac{{{d^2}\Psi }}{{d{\xi ^2}}} + \frac{{{P_m}}}{{\kappa {{(1 + {P_m})}^{1/2}}}}\frac{{{d^4}\Phi }}{{d{\xi ^4}}} \, .
\end{equation}
Following a standard procedure \cite{CGJ_1966,BondSobel84,PegSchep86,Porcelli_1987}, we introduce the Fourier transform of the perturbed quantities
\begin{equation} 
\left[ {\hat \Psi (\theta ),\hat \Phi \left( \theta  \right)} \right] = \int_{ - \infty }^{ + \infty } {\left[ {\Psi \left( \xi  \right),\Phi \left( \xi  \right)} \right]} \,{e^{ - i\theta \xi }}d\xi  \, , 
\end{equation}
which should be understood in a generalized sense \cite{Lighthill_1958}, since ${\Psi \left( \xi  \right)}$ and ${\Phi \left( \xi  \right)}$ are not square integrable. Then, the Fourier transformed Eqs. (\ref{inner_eq1}) and (\ref{inner_eq2}) can be combined to give the layer equation
\begin{equation}\label{layer_eq}
\begin{array}{l}
\dfrac{d}{{d\theta }}\left( {\dfrac{{{\theta ^2}}}{{\lambda  + {\kappa ^{ - 1}}{{\left( {1 + {P_m}} \right)}^{ - 1/2}}{\theta ^2}}}\dfrac{{d\hat \Phi }}{{d\theta }}} \right)  \\
\quad \quad \quad \quad \quad \quad \quad = \dfrac{\lambda }{{{\alpha ^2}}}{\theta ^2}\hat \Phi  + \dfrac{{{P_m}}}{{{\alpha ^2}\kappa {{\left( {1 + {P_m}} \right)}^{1/2}}}}{\theta ^4}\hat \Phi  \\
\end{array}
\end{equation}
The appropriate boundary condition that ensures the asymptotic matching to the outer solution must behave as 
\begin{equation}  \label{BC_theta_2}
\hat \Phi (\theta ) \to {\Phi _0}\left( {\frac{1}{\theta } + \frac{\pi }{{\Delta '}}\operatorname{sgn} \left( \theta  \right)} \right)  \quad \quad {\text{for}} \quad \theta  \to 0 \, ,
\end{equation}
where $\Phi_0$ is a constant. Furthermore, the layer equation must be solved subject to the condition 
\begin{equation}  \label{BC_theta_1}
\hat \Phi \left( \theta  \right) \to 0 \quad \quad {\text{for}}\quad \theta  \to \infty   \, . 
\end{equation}

In the limit $P_m \ll 1$, the eigenvalue problem (\ref{layer_eq})-(\ref{BC_theta_1}) yields the dispersion relation \cite{Coppi_1976, Lou_2007}
\begin{equation}  \label{disp_rel_Lou}
{\Lambda ^{5/4}}\frac{{\Gamma \left[ {\left( {{\Lambda ^{3/2}} - 1} \right)/4} \right]}}{{\Gamma \left[ {\left( {{\Lambda ^{3/2}} + 5} \right)/4} \right]}} =  - \frac{8}{\pi }{\left( {\kappa \alpha } \right)^{ - 1/3}}\Delta ' \, ,
\end{equation}
where $\Lambda  \equiv {\alpha ^{ - 2/3}}{\kappa ^{1/3}} \lambda$ and $\Gamma$ indicates the gamma function. This is the regime considered by Loureiro {\it et al.} \cite{Lou_2007}, who found the growth rate and the wavenumber of the fastest growing mode by balancing two relevant limits of Eq. (\ref{disp_rel_Lou}), namely 
\begin{equation}  \label{lim1}
\lambda = {\left[ {\frac{{\Gamma \left( {1/4} \right)}}{{2\pi \Gamma \left( {3/4} \right)}}} \right]^{4/5}}{\alpha ^{2/5}}{\kappa ^{ - 3/5}}{{\Delta '}^{4/5}}  \quad \quad {\text{for}} \quad \Lambda \ll 1 \, ,
\end{equation}
and 
\begin{equation}  \label{lim2}
\lambda = {\alpha ^{2/3}}{\kappa ^{ - 1/3}} - \frac{{2{\pi ^{1/2}}\alpha }}{{3\Delta '}}  \quad \quad {\text{for}} \quad \Lambda \to 1^- \, .
\end{equation}
In this way, they showed that for the fastest growing mode $\gamma_{\max} L_{cs}/v_{A,u} \sim S^{1/4}$ and $k_{\max } L_{cs} \sim S^{3/8}$.

Here, on the other hand, we are interested also in plasmas with non-negligible viscosity. In the case $P_m > 1$ an exact analytic solution of the layer equation (\ref{layer_eq}) has not been found, but we can rely on analytic approximations that are valid in different relevant asymptotic regimes \cite{Porcelli_1987}.

We first consider the small-$\Delta '$ regime, which is characterized by the condition $\Delta ' \delta_{\rm in} \ll 1$, with $\delta_{\rm in}$ indicating the inner layer width. In this case, Eq. (\ref{layer_eq}) can be solved by asymptotic matching the solutions obtained for large $\theta$ and relatively small $\theta$. For values of $\theta$ such that $\theta > (\lambda \kappa)^{1/2} P_m^{1/4}$, Eq. (\ref{layer_eq}) may be approximated as
\begin{equation}  \label{layer_eq_VR_1}
\frac{{{d^2}\hat \Phi_> }}{{d{\theta ^2}}} - \frac{{{\theta ^4}}}{{{\alpha ^2}{\kappa ^2}}}{\hat \Phi _ > } = 0 \, ,
\end{equation}
where $\hat \Phi_> (\theta)$ indicates the function that approximate $\hat \Phi (\theta)$ in the considered interval. The general solution of this equation is \cite{Comisso2015}
\begin{equation} 
\hat \Phi_> = {C_4}{\theta^{1/2}}{I_{1/6}}\left( {\frac{1}{3}} \theta^3 \delta_{\rm in}^3 \right) + {C_5}{\theta^{1/2}}{K_{1/6}}\left( {\frac{1}{3}} \theta^3 \delta_{\rm in}^3 \right)  \, ,
\end{equation}
where $I_{1/6}$ and $K_{1/6}$ are the modified Bessel functions of the first and second kind of order $1/6$, $C_4$ and $C_5$ are integration constants, and $\delta_{\rm in} = (\alpha \kappa)^{-1/3}$. On the other hand, if we consider values of $\theta$ such that $\theta < (\alpha \kappa)^{1/3}$, Eq. (\ref{layer_eq}) may be approximated as
\begin{equation}  \label{layer_eq_VR_5}
\frac{d}{{d\theta }}\left( {\frac{{{\theta ^2}}}{{\lambda  + {\kappa ^{ - 1}}P_m^{ - 1/2}{\theta ^2}}}\frac{{d\hat \Phi_< }}{{d\theta }}} \right) = 0  \, ,
\end{equation}
where $\hat \Phi_< (\theta)$ indicates the function that approximate $\hat \Phi (\theta)$ for relatively small values of $\theta$. The general solution of this equation is 
\begin{equation} 
{\hat \Phi _< } = {C_6} \left( { - \frac{\lambda}{\theta } + \kappa^{-1} {P_m^{-1/2}} \theta } \right) + {C_7}  \, ,
\end{equation}
where $C_6$ and $C_7$ are integration constants. Imposing the boundary conditions (\ref{BC_theta_2}) and (\ref{BC_theta_1}) and matching the functions $\hat \Phi_>$ and $\hat \Phi_<$ in the overlapping interval ${{(\lambda \kappa )^{1/2}} P_m^{1/4}}  < \theta  < {(\alpha \kappa )^{1/3}}$, we can finally obtain
\begin{equation}  \label{dispersion_1}
\lambda  = \frac{1}{{\pi {c_*}}}\frac{{{\alpha ^{1/3}}}}{{{\kappa ^{2/3}}P_m^{1/2}}} \Delta '  \, ,
\end{equation}
where
\begin{equation}
c_*  \equiv  {6^{2/3}}\frac{{\Gamma \left( 5/6 \right)}}{{\Gamma \left( 1/6 \right)}} \, .
\end{equation}

We now turn our attention to the large-$\Delta '$ regime ($\Delta ' \to \infty$). In this case, we can assume the ``visco-resistive'' ordering, for which all terms in Eq. (\ref{layer_eq}) are comparable over the entire width of the inner layer, with the exception of the inertial term, which is taken to be negligible \cite{Porcelli_1987}. With this ordering, estimating $\theta \sim 1/ \delta_{\rm in}$ and ${d}/d{\theta} \sim \delta_{\rm in}$, we obtain
\begin{equation}  \label{}
\delta_{\rm in} \sim (\alpha \kappa)^{-1/3}  \, 
\end{equation}
and %
\begin{equation}  \label{}
\lambda  \sim {\alpha^{2/3}} \kappa^{-1/3} P_m^{-1/2}  \, .
\end{equation}
A more precise multiplicative constant can be found through the numerical analysis of the layer equation (\ref{layer_eq}), which leads asymptotically (for large $P_m$ and $\Delta ' \to \infty$) to 
\begin{equation}  \label{dispersion_2}
\lambda  = c_{**} \frac{{{\alpha ^{2/3}}}}{{{\kappa ^{1/3}}P_m^{1/2}}}  \, ,
\end{equation}
where $c_{**} \approx 1.53$ \cite{Porcelli_1987}.

\subsection{Fastest growing mode} 

The most unstable mode, which is the one that dominates the plasmoids growth, occurs at the intersection of the small-$\Delta '$ and large-$\Delta '$ regimes. Therefore, by equating Eqs. (\ref{dispersion_1}) and (\ref{dispersion_2}), with the help of Eq. (\ref{Tearing_parameter}), we find the wavenumber of the fastest growing mode
\begin{equation}  \label{kappa_max}
\kappa_{\max} = k_{\max } L_{cs} \approx {\left( {\frac{2}{\pi}} \right)^{3/4}} \frac{{{\alpha ^{5/4}}{S^{3/8}}}}{{P_m^{3/16}}} \, .
\end{equation}
From this relation we have also the number of plasmoids that initially appear in the unstable current sheet, which is $N \approx \kappa_{\max} / 2\pi$. The $\lambda$-value of the fastest growing mode can be easily evaluated as
\begin{equation}  \label{lambda_max}
\lambda_{\max} = \frac{\gamma_{\max} L_{cs}}{v_{A,u} \kappa_{\max}} \approx \frac{3}{2}{\left( {\frac{\pi }{2}} \right)^{1/4}} \frac{{{\alpha ^{1/4}}}}{{{S^{1/8}}P_m^{7/16}}} \, ,
\end{equation}
which means that the growth rate of the plasmoid instability is
\begin{equation}  \label{gamma_max}
\gamma_{\max}  \frac{L_{cs}}{v_{A,u}} \approx \frac{3}{{\sqrt {2\pi} }} \frac{{{\alpha ^{3/2}}{S^{1/4}}}}{{P_m^{5/8}}} \, .
\end{equation}
Finally, the inner layer width of the fastest growing mode is
\begin{equation}  \label{delta_in_max}
\frac{\delta_{\rm in,\max}}{\delta_{cs}}  \approx {\left( {\frac{\pi }{2}} \right)^{1/4}} \frac{{P_m^{1/16}}}{{{\alpha ^{3/4}}{S^{1/8}}}} \, .
\end{equation}

Note that plasma viscosity has the effect of decreasing the growth rate and the wavenumber of the plasmoid instability, while it increases (slightly) the inner layer width. We observe that the same dependence of $\kappa_{\max}$, $\gamma_{\max}$ and $\delta_{\rm in,\max}$ on $S$ and $P_m$ was found in a heuristic way by Loureiro and coworkers \cite{LSU_2013}. 
Their scalings were obtained (non-rigorously) by applying the classical tearing mode expressions for large $P_m$ \cite{Porcelli_1987} to a Harris current sheet with equilibrium scale length $L \equiv \delta_{cs} \approx L_{cs} S^{-1/2} P_m^{1/4}$ \cite{Park_1984}, in analogy to what was done by Bhattacharjee {\it et al.} \cite{BHYR_2009} to reproduce the scaling relations derived by Loureiro {\it et al.} \cite{LSU_2013} for plasmas with $P_m \ll 1$. Here, instead, we have adopted a more rigorous approach that confirms the $S$ and $P_m$ dependencies and gives us the remaining multiplicative factors. Moreover, we observe that since the relations (\ref{kappa_max})-(\ref{delta_in_max}) for high-$P_m$ plasmas have the same $\alpha$ factors and $S$ dependencies of those resulting from Eqs. (\ref{lim1})-(\ref{lim2}) for low-$P_m$ plasmas, general relations for arbitrary $P_m$ can be represented as
\begin{equation}  \label{REL1}
{k_{\max }}{L_{cs}} \sim \frac{{{\alpha ^{5/4}}{S^{3/8}}}}{{{{\left( {1 + {P_m}} \right)}^{3/16}}}} \, ,
\end{equation}
\begin{equation}  \label{REL2}
\gamma_{\max}  \frac{L_{cs}}{v_{A,u}} \sim \frac{{{\alpha ^{3/2}}{S^{1/4}}}}{{{{\left( {1 + {P_m}} \right)}^{5/8}}}} \, ,
\end{equation}
\begin{equation}  \label{REL3}
\frac{\delta_{\rm in,\max}}{\delta_{cs}} \sim \frac{{{{\left( {1 + {P_m}} \right)}^{1/16}}}}{{{\alpha ^{3/4}}{S^{1/8}}}} \, .
\end{equation}

It is clear that, despite the damping effect of the plasma viscosity, the plasmoid instability turns out to be super-Alfv\'enic in plasmas of interest to this work, i.e., for very large $S$-values. 
Note that the growth rate of the plasmoid instability is super-Alfv\'enic with respect to the inverse Alfv\'en time $\tau_{A,L} = L_{cs}/v_{A,u}$ (along the current sheet), but it is slower than the inverse Alfv\'en time $\tau_{A,\delta} = \delta_{cs}/v_{A,u}$ (transverse to the current sheet), as is required for the tearing mode analysis. It should also be noted that while the decrease of the growth rate with increasing magnetic Prandtl number may occur as a dissipative effect of viscosity, it is also possible for a stable current sheet to be destabilized by the viscosity in analogy to what happens for the Orr-Sommerfeld equation in fluid dynamics. 
The plasma viscosity may indeed destabilize the stable branches of the dispersion relation that follows from Eqs. (\ref{eq7}) and (\ref{eq8}). For this reason, we believe that a treatment of the role of viscosity on the effects of marginal stability is warranted.

It is easy to verify that the obtained relations (\ref{REL1})-(\ref{REL3}) for the fastest growing mode justify the assumptions adopted to derive these relations, i.e., ${\kappa_{\max}^{-1}}{(1 + {P_m})^{-1/2}} \ll \lambda_{\max}  \ll 1$, $1/{\kappa_{\max}} \ll 1$, and ${\kappa_{\max}} \epsilon  \ll 1$ (recall that $\epsilon = S^{-1/2} {\left( {1 + {P_m}} \right)^{1/4}} \ll 1$ since we are considering highly elongated current sheets).

Furthermore, if we suppose that the stabilizing effect of the flow becomes ineffective for current sheets exceeding the critical aspect ratio $L_c/ \delta_c \equiv 1/ \epsilon_c$, it is straightforward to see that
\begin{equation} \label{Plasmoid_Condition_2}
S_c = \epsilon_c^{-2} {\left( 1 + P_m \right)^{1/2}} \, ,
\end{equation}
i.e., the critical Lundquist number for the stability of the sheet increases with increasing plasma viscosity due to the reduction of the outflow velocity \cite{Comisso2015,Comisso2015b}. In particular, in the limit $P_m \ll 1$ we get $S_c \approx \epsilon_c^{-2}$ \cite{BHYR_2009,Cassak_2009}, whereas in the limit $P_m \gg 1$ it follows that $S_c \approx \epsilon_c^{-2} P_m^{1/2}$, which agrees with the condition proposed by Loureiro {\it et al.} \cite{LSU_2013}.

\section{Nonlinear Theory} \label{sec4}

The previous linear analysis breaks down and nonlinearities must be considered when the plasmoid chain width $w$ grows to a size comparable or exceeding the linear layer width $\delta_{\rm in,\max}$. Then, in order to determine the nonlinear evolution of the plasmoids, we have to consider the relevant nonlinear regime that characterizes their growth.

\subsection{Nonlinear evolution}

To ascertain the proper nonlinear regime, we observe that for the fastest growing mode
\begin{equation} \label{}
\Delta '_{\max} \delta_{cs} \approx \frac{{2{\alpha ^2}}}{{\kappa_{\max} \epsilon }} \sim \frac{{2{\alpha ^{3/4}}{S^{1/8}}}}{{{{\left( {1 + {P_m}} \right)}^{1/16}}}} \, ,
\end{equation}
while the plasmoid half-width is $w \sim {\delta _{{\rm{in}},\max }}$ at the beginning of the nonlinear evolution. Therefore, when the fastest growing mode enters into the nonlinear regime
\begin{equation} \label{}
\Delta '_{\max} w \sim 1 \, 
\end{equation}
(ignoring a factor 2), which implies that the plasmoids evolve nonlinearly according to a Waelbroeck regime \cite{Waelb1993} ($\Delta '_{\max} w \gtrsim 1$) instead of a Rutherford one \cite{Rutherford_1973} ($\Delta '_{\max} w \ll 1$). This means that the inter-plasmoids $X$-points collapse to form thin inter-plasmoids current sheets soon after entering the nonlinear regime.

The nonlinear growth of the plasmoids can be determined by matching a model of magnetic reconnection within current sheets with Waelbroeck's solution for the magnetic configuration of rapidly reconnecting islands \cite{Waelb1989}. 
In this case, the temporal rate of change of magnetic flux at the separatrix of a plasmoid can be evaluated as
\begin{equation}  \label{}
\frac{{d{\psi_s}}}{{dt}}  \approx  \eta \frac{B_{y}^*}{\delta_{cs}^*}  \, ,
\end{equation}
where $B_{y}^*$ is the magnetic field upstream of the inter-plasmoids current sheet and $\delta_{cs}^*$ is the half-width of this sheet.
Considering Eq. (\ref{delta_cs}) for the case of the secondary reconnecting current sheet, we can estimate $\delta_{cs}^*$ to obtain \cite{Park_1984} 
\begin{equation}  \label{Rec_rate_at_X_3}
\frac{{d{\psi_s}}}{{dt}}  \approx   \frac{{{\eta ^{1/2}}}}{{{{\left( {1 + {P_m}} \right)}^{1/4}}}} {\left( {\frac{{B_y^{*3}}}{{L_{cs}^*}}} \right)^{1/2}}  \, .
\end{equation}

The time scale of the magnetic flux change across the inter-plasmoid layer is 
\begin{equation}  \label{}
\tau _\psi = \frac{{{\psi_s}}}{{d{\psi_s}/dt}} \approx \frac{w}{{\eta \Delta '_{\max} }}  \, ,
\end{equation}
which is much larger than the transverse Alfv\'en time $\tau_{A,\delta} = \delta_{cs}/v_{A,u}$. Hence, outside the singular layer, the magnetic configuration proceeds through a sequence of MHD equilibria. Furthermore, in this regime $\tau _\psi$ is smaller than the characteristic time for current diffusion, the plasmoid skin time $\tau_w = w^2/ \eta$. This implies that the flux becomes frozen-in and the magnetic helicity is conserved within pairs of flux tubes \cite{Kadomtsev1975,Waelb1989}. 
We can therefore consider Waelbroeck's solution for rapidly reconnecting islands \cite{Waelb1989,Zakharov1993}, which tells us that the plasmoids become self-similar as $w$ grows for $w \gg 1/ \Delta '_{\max}$ (and $w < \delta_{cs}$), with a plasmoid magnetic configuration given by
\begin{equation}  \label{}
w \sim {\left( {\frac{{{\psi_s}}}{\alpha }} \right)^{1/2}}  , \quad  B_{y}^* \sim \alpha w \, , \quad  L_{cs}^*  \sim  \frac{{1}}{k_{\max }} \, .
\end{equation}

Substituting these relations into Eq. (\ref{Rec_rate_at_X_3}), we obtain the {\it plasmoid width evolution equation}
\begin{equation} \label{}
\frac{dw}{dt} \sim \frac{1}{2} {\left[ {\frac{ \eta \alpha }{{{(1 + P_m)}^{1/2}}} {k_{\max }}w} \right]^{1/2}} \, ,
\end{equation}
which yields the algebraic growth law
\begin{equation} \label{Nonlinear_growth}
w \sim \frac{\alpha}{16}\frac{ \eta {k_{\max}}}{{(1 + {P_m})}^{1/2}} {t^2} \, .
\end{equation}
Therefore, the growth of the plasmoids slows down from the exponential growth of the linear stage to an algebraic (quadratic) growth in time. Since $k_{\max}$ is given by Eq. (\ref{REL1}), after restoring the dimensions, one can obtain the relation
\begin{equation} \label{Nonlinear_growth_2}
\frac{w}{\delta_{cs}} \sim \frac{\alpha^{9/4}}{16}\frac{S^{3/8}}{{{\left( {1 + {P_m}} \right)}^{19/16}}}{\left( {\frac{t}{\tau_{A,L}}} \right)^2}  \, ,
\end{equation}
which tells us how the nonlinear evolution of the plasmoids depends on the Lundquist and magnetic Prandtl numbers. In the case of negligible plasma viscosity, this relation reduces simply to $w / \delta_{cs}  \sim \frac{\alpha^{9/4}}{16}  {S^{3/8}} {\left( {t/{\tau_{A,L}}} \right)^2} $.

It is important to point out that since $\gamma_{\max} \tau_{A,L} \propto S^{1/4}$ and ${\delta_{\rm in,\max}}/{\delta_{cs}} \propto S^{-1/8}$, in high-Lundquist-number systems the plasmoids spend an extremely short period of time in the linear regime. It is therefore the nonlinear regime that sets out the time scale for the disruption of the current sheet. From Eq. (\ref{Nonlinear_growth_2}) we can see that $w \sim \delta_{cs}$ on the time scale
\begin{equation} \label{}
\tau_{NL} \sim S^{-3/16} {(1 + {P_m})^{19/32}}{\tau _{A,L}}  \, .
\end{equation}
Therefore, the time scale for the current sheet disruption decreases with increasing Lundquist number but increases with increasing magnetic Prandtl number.

It should be noted that for extremely large $S$-values also the secondary current sheets connecting the plasmoids may  themselves be subject to the plasmoid instability if
\begin{equation}\label{}
\begin{array}{l}
\dfrac{{\delta _{cs}^*}}{{L_{cs}^*}} = {(1 + {P_m})^{1/4}}{\left( {\dfrac{{L_{cs}^*v_{A,u}^*}}{\eta }} \right)^{ - 1/2}}  \\
\quad \quad \quad \quad \quad \;  \sim {\left[ \dfrac{\delta_{cs}}{w} {\dfrac{{(1 + {P_m})}^{5/16}}{S^{5/8}}} \right]^{1/2}} < \epsilon_c \,. \\
\end{array}
\end{equation}
The same reasoning can be applied to the tertiary current sheets connecting the secondary plasmoids, and so forth.
This fractal-like cascade process \cite{Shibata_2001} towards smaller scales ends when the $(n)$th current sheets are sufficiently thick to avoid the plasmoid instability. Therefore, using $S^{(n)} \sim S_c$ one can find that, after the linear growth, the plasmoids separated by the $(n)$th current sheets destroy the $(n-1)$th current sheets on the nonlinear time-scale 
\begin{equation} \label{}
\tau_{NL}^{(n)} \sim \epsilon_c^{3/8} {(1 + {P_m})^{1/2}} \tau_{A,L}^{(n)} \, .
\end{equation}
This sets off a ``domino effect'' that speeds up the disruption of the global current sheet, which, in turn, allows for a very rapid release of magnetic energy.

\subsection{Implications for fast magnetic reconnection}

Plasmoids have a great impact on the global reconnection rate if they grow to a size $w \gtrsim \delta_{cs}$ before being expelled from the global current sheet. In this case, the global current sheet breaks up and is replaced by a chain of plasmoids of different sizes separated by smaller current sheets \cite{Shibata_2001,BHYR_2009,ULS_2010,Barta_2011,Loureiro2012}. 
In a plasmoid-dominated reconnection layer the dynamics can be particularly complex, with plasmoids constantly being generated, ejected and merging with each other. This leads to a strongly time-dependent reconnection process. However, we may expect this process to reach a statistical steady state with a marginally stable current sheet located at the main $X$-point. 
The statistical steady state expectation is supported by many numerical simulations from different research groups \cite{BHYR_2009,Cassak_2009,HB_2010,Fermo2011,Loureiro2012,Ni_2012,HB_2012,Ni_2013,Sironi2014,CGW_2014,Comisso2015}. 
The supposition of a marginally stable current sheet at the main $X$-point is also well-founded. Indeed, the fractal-like cascade process towards smaller scales due to the plasmoid instability ends when the length of the final local current layer is shorter that the critical length $L_c$. 
Also, the local current sheet at the main $X$-point is continuously stretched by the plasmoids moving in the outflow direction, thereby being regularly subject to the plasmoid instability every time its length exceeds $L_c$. Hence, it is reasonable to expect that the length of the local current sheet at the main $X$-point should always be quite close to $L_c$.

We can evaluate the global reconnection rate in the plasmoid-dominated regime as the rate of change of the magnetic flux reconnected at the main $X$-point \cite{ULS_2010}. This is because only the open-flux parcels matter when calculating the total reconnection rate, and summing and subtracting the contributions of all the reconnection layers one can find that the global reconnection rate is given only by the reconnection region at the main $X$-point. Therefore, the reconnection rate in statistical steady state can be calculated as
\begin{equation}  \label{Rec_rate_plasmoids_at_X_1}
\left\langle {{{\left. {\frac{{d\psi }}{{dt}}} \right|}_X}} \right\rangle  \approx \eta \frac{{{B_{y}}}}{{{\delta _c}}}  \, ,
\end{equation}
where $\left\langle  \ldots  \right\rangle $ denotes time-average and 
\begin{equation}  \label{delta_c}
\delta_c = \epsilon_c L_c = {\left( {\eta \frac{L_{c}}{v_d}} \right)^{1/2}}  \, .
\end{equation}
Using the global reconnecting magnetic field $B_u$ as an estimation of $B_y$ upstream of the current sheet at the main $X$-point, it follows that the local downstream velocity can be written as $v_d = v_{A,u} {\left( {1 + P_m} \right)^{-1/2}}$. 
Furtermore, using Eq. (\ref{Plasmoid_Condition_2}) for the critical Lundquist number, the reconnection rate in statistical steady state becomes simply
\begin{equation} \label{Rec_rate_plasmoids_at_X_2}
\left\langle {{{\left. {\frac{{d\psi }}{{dt}}} \right|}_X}} \right\rangle \approx \epsilon_c \frac{{v_{A,u}}{B_{u}}}{{{{\left( {1+P_m} \right)}^{1/2}}}}  \, .
\end{equation}

Note that numerical simulations  \cite{Bisk_1986,Lou_2005,BHYR_2009,SLUSC_2009,Cassak_2009,LCV_2010,SC_2010,Skender_2010,HB_2010,Loureiro2012,Baty_2012} indicate $1/\epsilon_c \sim 10^2$, implying that the plasmoid-dominated regime is a fast reconnection regime provided that $P_m$ is not too high, as it could be in some astrophysical environments like the warm interstellar medium and protogalactic plasmas \cite{BrandSubram2005}.

Expression (\ref{Rec_rate_plasmoids_at_X_2}), first obtained in Ref. \cite{Comisso2015}, extends the expressions valid for $P_m \ll 1$ ($\left\langle {{{\left. {d\psi /dt} \right|}_X}} \right\rangle  \approx  \epsilon_c v_{A,u} B_{u}$) \cite{HB_2010,ULS_2010} and $P_m \gg 1$ ($\left\langle {{{\left. {d\psi /dt} \right|}_X}} \right\rangle  \approx  \epsilon_c v_{A,u} B_{u} P_m^{-1/2}$) \cite{CGW_2014} to arbitrary magnetic Prandtl numbers. Note that the statistical steady-state reconnection rate in the plasmoid-dominated regime becomes nearly independent of the microscopic plasma parameters only for low magnetic Prandtl numbers. For large magnetic Prandtl numbers this behavior changes because the viscous energy dissipation leads to a decrease of the outflow velocity.

Finally, it is important to point out that the plasmoid instability may also lead to an even faster reconnection regime by triggering a transition to collisionless reconnection \cite{Daugh_2009,SC_2010,ULS_2010,HBS_2011,Ji_Daughton_2011}. Indeed, the fractal-like cascade process caused by the plasmoid istability produces local reconnection layers that may be in the collisionless regime even if the initial global current sheet is collisional, i.e., even if $\delta_{cs} \gg l_k$, being $l_k$ a relevant kinetic length scale. This transition occurs if $\delta_{c} \lesssim l_k$. Therefore, by rewriting Eq. (\ref{delta_c}) as ${\delta _c} \approx {L_{cs}}{(1 + {P_m})^{1/4}} (S_c/S^2)^{1/2}$, we find that the plasmoid instability leads to the collisionless regime if 
\begin{equation} \label{}
S \gtrsim  \frac{L_{cs}}{\epsilon_c l_k} {(1 + {P_m})^{1/2}}  \, .
\end{equation}
This relation exhibits a significant dependence on the magnetic Prandtl number, which has the effect of increasing the value of the global Lundquist number required to reach the collisionless regime. 
Note also that the relevant kinetic length scale $l_k$ depends on the value of the out-of-plane magnetic field. 
In the case of negligible guide magnetic field $l_k=d_i$ (see, e.g., \cite{MB_1996,CSD_2005,SimaChac2008}), being $d_i=c / \omega_{pi}$ the ion skin depth, while in the opposite case of strong guide magnetic field $l_k = \rho_\tau$ (see, e.g., \cite{KDW_1995,CDS_2007,Comisso2013}), where $\rho_\tau = c_s / \omega_{ci}$ represents the ion sound Larmor radius based on both the electron and ion temperatures.

\section{Conclusions} \label{sec5}

In this paper we have presented a theory of the plasmoid instability in visco-resistive plasmas, i.e., plasmas in which both resistivity and viscosity are important. We have considered both the linear and nonlinear regimes of the plasmoid instability, and we have also evaluated its effects on the global reconnection rate.

The linear analysis presented here generalizes the theoretical work performed by Loureiro {\it et al.} \cite{Lou_2007} to account for non-negligible values of plasma viscosity.
The linear growth rate and the wavenumber of the visco-resistive plasmoid instability are found to be $\gamma_{\max} {\tau_{A,L}} \sim \alpha^{3/2} S^{1/4} {\left( {1 + {P_m}} \right)}^{-5/8} $ and ${k_{\max }}{L_{cs}} \sim \alpha^{5/4} S^{3/8} {\left( {1 + {P_m}} \right)}^{-3/16}$, respectively. Therefore, plasma viscosity has the effect of decreasing the linear growth rate and the wavenumber of the instability. However, despite its damping effect, for very high Lundquist numbers the plasmoid instability turns out to be very rapid (super-Alfv\'enic) in the linear regime.

Nonlinearities begin to be important when the width of the plasmoids becomes of the order of the linear layer width, which is found to be ${\delta _{{\rm{in}},\max }} / \delta_{cs} \sim  \alpha^{-3/4} S^{-1/8} {{{{\left( {1 + {P_m}} \right)}^{1/16}}}}$. Since ${\delta _{{\rm{in}},\max }}/\delta_{cs}$ decreases with increasing Lundquist number, the nonlinear growth of the plasmoids turns out to be fundamental in setting out the time-scale for the disruption of high-$S$ current sheets. 
For $w \gg 1/ \Delta '_{\max}$ (soon after the beginning of the nonlinear regime) we have found that the half-width of the plasmoids grows as $w/\delta_{cs} \sim \frac{\alpha^{9/4}}{16} {S^{3/8}}{\left( {1+P_m} \right)^{-19/16}}{\left( {t/{\tau_{A,L}}} \right)^2}$. Therefore, the growth of the plasmoids slows down from the exponential growth of the linear regime to an algebraic growth characterized by a time scale $\tau_{NL} \sim S^{-3/16} {(1 + P_m)^{19/32}}{\tau _{A,L}}$.
For extremely large Lundquist numbers, a fractal-like current sheet structure with hierarchical plasmoid chains can speed up the growth of the larger plasmoids and consequently also the destruction of the global current sheet.

Finally, we have shown that also in visco-resistive plasmas the plasmoid instability is pivotal in allowing fast magnetic reconnection. In particular, we have calculated that the reconnection rate in statistical steady-state is $\left\langle {{{\left. {d\psi /dt} \right|}_X}} \right\rangle  \approx  \epsilon_c v_{A,u} B_{u} {(1 + {P_m})^{-1/2}}$, independent of the Lundquist number but not on the magnetic Prandtl number. For not excessively high $P_m$-values, the reconnection rate is fast, i.e., a significant fraction of the out-of-plane electric field evaluated upstream of the global reconnection layer. We have also shown that the plasmoid instability may allow fast reconnection by leading to a collisionless reconnection regime if $ S \gtrsim L_{cs} {(\epsilon_c l_k)^{-1}} {(1 + {P_m})^{1/2}}$.

\acknowledgments

We benefited from stimulating discussions with Amitava Bhattacharjee, Yi-Min Huang, Manasvi Lingam, Nuno F. Loureiro and Fran\c{c}ois L. Waelbroeck. 
We acknowledge financial support from the European Community under the contracts of Association between Euratom and ENEA.

\end{document}